\documentclass[10pt,journal,epsfig,twoside]{IEEEtran}
\usepackage{amsmath}
\usepackage{graphicx}
\usepackage{amssymb}
\usepackage{amsfonts}
\usepackage{cite}
\usepackage{bm,comment}
\usepackage{algorithm}
\usepackage{algpseudocode}

\usepackage{subeqnarray}
\usepackage{subfigure}
\usepackage{multirow}
\usepackage{slashbox}
\usepackage{stfloats}
\usepackage{float}
\newenvironment{sequation}{\begin{equation}\small}{\end{equation}}

\newtheorem{proposition}{Proposition}
\newtheorem{remark}{Remark}

\newtheorem{lemma}{Lemma}
\newtheorem{corollary}{Corollary}

\begin{document}
\IEEEoverridecommandlockouts
\title{Channel Estimation and Transmission for Intelligent Reflecting Surface Assisted THz Communications}
\author{\IEEEauthorblockN{Boyu Ning\IEEEauthorrefmark{1}, Zhi Chen\IEEEauthorrefmark{1}, Wenrong Chen\IEEEauthorrefmark{1}, and Yiming Du\IEEEauthorrefmark{1}}\\
\IEEEauthorblockA{\IEEEauthorrefmark{1}National Key Laboratory of Science and Technology on Communications\\
University of Electronic Science and Technology of China (UESTC), Chengdu 611731, China\\
Emails: boydning@outlook.com; chenzhi@uestc.edu.cn; wenrongchen@std.uestc.edu.cn;}
\thanks{This work was supported by the National Key R$\&$D Program of China under Grant 2018YFB1801500.}}
\maketitle

\begin{abstract}
Intelligent reflecting surface (IRS) is envisioned as a promising technology to broaden signal coverage and enhance transmission in terahertz (THz) communications. Due to the passivity of IRS, the channel measurement can not be achieved by traditional pilot manner and the subsequent cooperative transmission design remains an open problem. This paper investigates the channel estimation and transmission solutions for massive multiple input multiple output (MIMO) IRS-assisted THz system. The channel estimation is realized by beam training and the quantization error is analyzed for evaluating performance. In addition, a novel hierarchical search codebook design is proposed as a low-complexity basis of beam training. Based on above foundations, we propose a cooperative channel estimation procedure to tactfully acquire the channel knowledge. Finally, by leveraging obtained channel information, the designs of IRS and transceivers are directly provided in closed form without reconstructing the full channel matrix or additional optimization. Simulation and numerical results are presented to illustrate the minimum signal to noise ratio (SNR) required for beam training and the efficacy of the proposed transmission solutions.
\end{abstract}

\IEEEpeerreviewmaketitle

\section{Introduction}
As a promising technology to support explosive growth of data traffic, terahertz (THz) communication emerges as a key candidate in future generations\cite{IF}. Due to the severe signal attenuation and narrow wave spread in THz frequencies (0.1T-10T), the number of effective propagation paths is quite limited and the THz channel is sparse with line-of-sight (LoS) dominant\cite{Track,Jornet}. Therefore, the transmission is extremely sensitive to LoS-cut obstacles and the enhancement of signal coverage for THz communication remains a critical problem.

Recently, an emerging hardware technology for broadening signal coverage with reduced energy consumption and low-cost implementation is the so-called intelligent reflecting surface (IRS)\cite{ZL}. What makes the IRS attractive compared to fixed reflector is the possibility to induce a certain phase shift independently on the incident electromagnetic by controllable meta-material\cite{cometa,gain}. Compared to conventional relay schemes, IRS-assisted schemes enhance the transmission without receiving or transmitting signals, however, by altering the propagation environments. Hence, with the emergence of IRS, the  cooperative transmission design for IRS-assisted schemes has attracted much attention and concerns from researchers. Prior IRS-related works \cite{huangchi,qinte2,ism,ghy,nby,huangchi2} mainly aim to maximize the spectral efficiency by optimizing the IRS and the transmitter beamformer for multi-input single-output (MISO) systems under perfect channel state information (CSI). However, these solutions can not be applied to IRS-assisted THz communication due to the following concerns.
\begin{itemize}
\item  To compensate severe signal attenuation in THz communication, massive
multiple input multiple output (MIMO) implementation needs to be considered to provide enough array gain. The algorithms in \cite{huangchi,qinte2,ism,ghy,nby,huangchi2} for cooperative optimization are oriented for single antenna receiver, and can not be extended to MIMO systems.

\item Traditional fully digital beamformer (FDB) requires one dedicated radio-frequency (RF) chain for each antenna in the MIMO systems, which results in huge energy consumption and unaffordable hardware cost. Thus, an appealing low-complexity hybrid beamforming architecture should be retorted as an alternative\cite{5-2,5,jingdian}.

\item Owing to the passivity of IRS (unable to send and receive signals), the CSI of IRS-assisted system is inaccessible by traditional channel estimation approaches\cite{csi1,csi1a,csi3}. As a result, the acquisition of the overall channel information in IRS-assisted systems becomes the mandatory and primary issue. Meanwhile, the cooperative transmission solution should be derived based on the form of obtained channel knowledge.  
\end{itemize}
For above-mentioned reasons, three state-of-the-art technologies, i.e., THz communication, IRS-assisted transmission, and massive MIMO techniques are possible and necessary to be integrated in wireless system for expected orders of magnitude gain. Thus, the development of such a triad-specific transmission scheme is imperative and significant.

In this paper, we investigate the channel estimation and transmission solutions for THz massive MIMO IRS-assisted system with hybrid beamforming architectures. By exploiting the sparsity of THz channel and the characteristics of massive antenna array, a beam training manner is proposed as the basis for realizing channel measurement, and the beam patterns (or beam directions) distribution as well as the quantization error caused by beam training are analyzed and evaluated. To reduce the complexity of the channel estimation, a novel hierarchical search codebook design is proposed, in which beams in the same stage are homogeneous in different directions. Based on the above foundations, a cooperative estimation procedure is proposed to measure the channel of IRS-assisted scenarios. Specifically, the deaf-mute IRS carries out a set of prescribed operations step by step in estimation phase, which are known to active transceivers. By leveraging the prescribed prior information, corresponding channel knowledges can be obtained by active terminals and sent to IRS controller. Then, the IRS design is provided to optimally bridge the transmission links, in which the effective channel is further estimated by beam training. Without CSI matrix reconstruction, the hybrid precoder/combiner and IRS designs are straightforwardly provided in closed form via obtained channel knowledge. Simulations results are presented to illustrate the minimum signal to noise ratio (SNR) required for beam training and numerical results indicate that the spectral efficiencies achieved by the proposed designs significantly outperform non-IRS-assisted benchmark and approach the rate achieved by optimal fully digital precoder/combiner with perfect CSI.
\section{System Model}
\subsection{Massive MIMO with Hybrid Architectures}
\begin{figure}[t]
\centering
\includegraphics[width=3.5in]{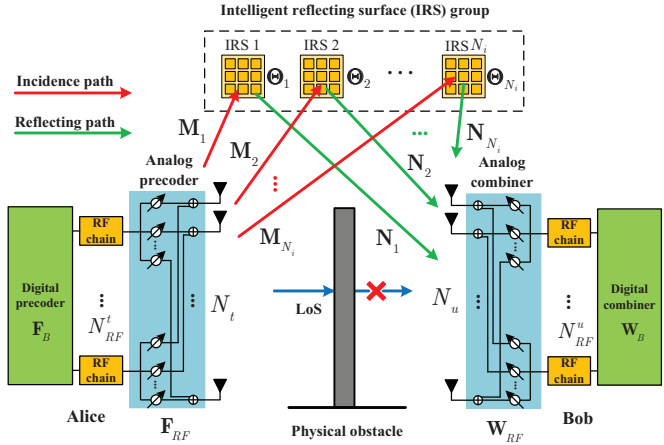}
\caption{A massive MIMO IRS-assisted system with the hybrid beamforming architecture at Alice and Bob terminal.}\label{figmimo}
\vspace{-12pt}
\end{figure}
As depicted in Fig. \ref{figmimo}, we consider a narrowband downlink massive MIMO system  where Alice with $N_t$ antennas intends to send $N_s$ data streams to Bob with $N_u$ antennas. Since the implementation of fully digital precoder/combiner suffers from high cost and power consumption, we consider hybrid digital and analog architecture. $N_{RF}^t$ RF chains and $N_{RF}^u$ RF chains are implemented at Alice and Bob such that $N_s \leq {N_{RF}^t} \ll {N_t}$ and $N_s \leq {N_{RF}^u} \ll {N_u}$.

At Alice with hybrid precoder, a symbol vector ${\bf{s}} \in {C^{{N_s}}}$ is first processed by a digital precoder ${{\bf{F}}_{B}} \in {\mathbb{C}^{{{N_{RF}^t} \times {N_s}}}}$, and then up-converted to analog procoder ${{\bf{F}}_{RF}} \in {\mathbb{C}^{{N_t} \times {N_{RF}^t}}}$. The analog beamformer is implemented by variable phase shifters and subjected to magnitude constraint, i.e., $\left| {{{\bf{F}}_{RF}}(i,j)} \right| = 1,\;\forall i,j$. Here, we assume that $E[{\bf{s}}{{\bf{s}}^H}] = {{\bf{I}}_{{N_s}}}$. Given the normalized power constraint  ${\| {{{\bf{F}}_{RF}}{{\bf{F}}_{B}}} \|_2^2} = 1$ and total transmit power $P$, the received signal can be expressed as
\begin{equation}
{\bf{y}} = \sqrt P {\bf{H}}{{\bf{F}}_{RF}}{{\bf{F}}_B}{\bf{s}} + {\bf{n}},
\end{equation}
where ${{\bf{H}}} \in {{\mathbb{C}}^{{N_u} \times {N_t}}}$ represents the overall channel  matrix between Alice and Bob. ${\bf{n}} \in {\mathbb{C}^{{N_u}}}$ are zero-mean additive white Gaussian noise, i.e., ${\bf{n}} \sim \mathcal{CN}(\mathbf{0},\sigma _n^2\mathbf{I}_{N_u})$. The received signal ${\bf{y}} \in {\mathbb{C}}^{{N_u}}$ is first processed by an analog combiner ${\bf{W}}_{RF}^k \in {\mathbb{C}}^{{N_u} \times {N_{RF}^u}}$ such that $\left| {{\bf{W}}_{RF}^k(i,j)} \right| = 1,\;\forall i,j$, and then down-converted to digital combiner ${\bf{W}}_{B}^k \in {\mathbb{C}}^{{N_{RF}^u} \times {N_q}}$, which results the final signals in 
\begin{equation}\label{shoufa}
{\bf{y}} = \sqrt P {\bf{W}}_B^H{\bf{W}}_{RF}^H{{\bf{H}}}{{\bf{F}}_{RF}}{{\bf{F}}_B}{\bf{s}} + {\bf{W}}_B^H{\bf{W}}_{RF}^H{\bf{n}}.
\end{equation}
The overall spectral efficiency is given as \cite{jingdian}
\begin{equation}\label{se}
\begin{split}
R &= {\log _2}{\rm{det}}\Big( {{{\bf{I}}_{{N_s}}} + P{\bf{C}}^{-1}{\bf{W}}_B^H{\bf{W}}_{RF}^H{{\bf{H}}_k}{{\bf{F}}_{RF}}{{\bf{F}}_B}} \\
& {\qquad\qquad\qquad\;\;\;\;\;\;\;\;\;\;\;\;\;\;\;\; \times {\bf{F}}_B^H{\bf{F}}_{RF}^H{\bf{H}}_k^H{{\bf{W}}_{RF}}{{\bf{W}}_B}} \Big),
\end{split}
\end{equation}
where ${\bf{C}}=\sigma _n^2{\bf{W}}_B^H{\bf{W}}_{RF}^H{{\bf{W}}_{RF}}{{\bf{W}}_B}$ is the noise covariance. We mention that the uplink expression of considered system is identical to (\ref{shoufa}) with the roles of the precoder and combiner switched under transposed reciprocal channel, i.e., ${\bf{H}} \to {{\bf{H}}^T}$.
\subsection{IRS Assisted THz Channel Model} 
The large-scale antenna arrays and high pathloss at THz frequencies significantly limit scattering. Thus, the signal coverage is LoS dominant and is easily blocked by obstacles in THz communication\cite{IF,Jornet}. As shown in Fig. \ref{figmimo}, this paper considers such an indoor scenario that LoS path is blocked and $N_i$ IRSs ($\le N_{RF}^t$) are installed on surrounding wall to assist communication by providing tunable strong reflecting component.
Since the scattering components (reflected by other surfaces) are negligible, the overall channel ${{\bf{H}}}$ can be further expressed as
\begin{equation}\label{chan}
{{\bf{H}}}(f,{{\bf{d}}})\, = \sum\limits_{l = 1}^{N_i} {{\bf{H}}_{\rm{Ref}}(f,{d_{l}})},
\end{equation}
in which $\{{\bf{H}}_{\rm{Ref}}(f, {d_{l}})\} _{l = 1}^{N_i}$ are the strong reflecting channels provided by IRSs. $f$ is the carrier frequency and ${{\bf{d}}}=[d_{1},d_2, ... , d_{N_i}]$ is the vector of transmit distance in corresponding propagation paths. Each IRS can dynamically adjust $N_r$ intelligent elements of the meta-surface to achieve phase shifts on the reflecting signals by a passive way. Therefore, the received signal vector at the $l$th IRS can be linearly transformed by a diagonal phase-shift matrix ${\bf{\Theta }}_l = {\rm{diag}}(\beta {e^{j{\theta _1^l}}},\beta {e^{j{\theta _2^l}}}, \cdots ,\beta {e^{j{\theta _{{N_r}}^l}}}),\;l=1,...,N_i,$ and reflected to Bob, where $j = \sqrt { - 1}$, $\{ {\theta _i^l}\} _{i = 1}^{{N_r}} \in [0,2\pi )$, and $\beta \in [0,1]$ are imaginary unit, PSs, and amplitude reflection coefficient on the combined incident signal, respectively. Thus, the $N_i$ strong reflecting channels in (\ref{chan}) can be expressed as
\begin{equation}\label{chan2}
\sum\limits_{l = 1}^{{N_i}} {{{\bf{H}}_{{\rm{Ref}}}}(f,{d_l})}  = \sum\limits_{l = 1}^{{N_i}} {{\eta _l}G_t G_r{{\bf{N}}_l}(f,d_l^N){{\bf{\Theta }}_l}} {{\bf{M}}_l}(f,d_l^M),
\end{equation}
where $\eta_l$ is the path-loss compensation factor (see{\ref{comp}}). $G_t$ and $G_r$ are the transmit and receive antenna gains. ${{\bf{M}}_l}$ is the Alice-IRS$l$ channel, and ${{\bf{N}}_l}$ is the IRS$l$-Bob channel. $d_l^M$ (resp. $d_{l}^N$) is the distance between Alice and $l$th IRS (resp. $l$th IRS and Bob) and $d_{l}=d_l^M+d_{l}^N$. The area of each IRS is limited and only provide one strong reflecting propagation path. Thus, the separate components in (\ref{chan2}) can be explicitly written as\cite{Track}
\begin{equation}\label{sepchan}
\begin{split}
{{\bf{M}}_l}(f,d_l^M) &= a(f,d_l^M){\bf{a}}_{{N_r}}^R\left( {\varphi _{R,M}^l} \right){\bf{a}}_{{N_t}}^A{\left( {\varphi _{A,M}^l} \right)^H},\\
{\bf{N}}_l(f,d_{l}^N) &= a(f,d_{l}^N){\bf{a}}_{{N_u}}^B\left( {\varphi _{B,N}^{l}} \right){\bf{a}}_{{N_r}}^R{\left( {\varphi _{R,N}^{l}} \right)^H},\\
\end{split}
\end{equation}
where $l = 1,...,{N_i}$, and $a(f,d)$ is the path loss which is drastically affected by the molecular absorption. Thus, $a(f,d)$ consists of the free-spread loss and the molecular absorption loss as\cite{Jornet}
\begin{equation}\label{loss}
a(f,d) = {\frac{c}{{4\pi fd}}}{e^{-\frac{1}{2}\tau (f)d}},
\end{equation}
where $c$ stands for the speed of light and $\tau (f)$ is the medium absorption factor. According to the path loss of the far-field RIS-assisted beamforming case\cite{gain}, the cascade path loss of BS-RIS$l$-user$k$ link is supposed to satisfy 
\begin{equation}
{G_t}{G_{\rm{r}}}{\eta }a(f,d_l^M)a(f,d_{k,l}^N) = \frac{{{G_t}{G_r}G{N_r}c}}{{8\sqrt {{\pi ^3}} fd_l^Md_{k,l}^N}}{e^{ - \frac{1}{2}\tau (f)(d_l^M + d_{k,l}^N)}}.
\end{equation}
where $G$ is the RIS element gain. Thus, the path-loss compensation factor is given as
\begin{equation}\label{comp}
{\eta } = \frac{{2\sqrt \pi  fG{N_r}}}{c}.
\end{equation}
In (\ref{sepchan}), the variable ${\varphi}$ is the path's azimuth angles of departure and arrival (AoD/AoA)\footnote{For explicit ${\varphi}$, the subscript A, R, and B represent Alice, IRS, and Bob; and the subscript H, M, and N represent the channel ${\bf{H}}$, ${\bf{M}}$, and ${\bf{N}}$ respectively. The superscript $l$ represents the $l$th IRS.}. In addition, ${{\bf{a}}_{N_a}^E}(\phi)$ is the normalized antenna array response vectors at terminal $E$ with $N_a$ antenna elements. For simplicity of exposition, (\ref{sepchan}) considers a uniform linear array (ULA) configuration such as 
\begin{equation}\label{ula}
{{\bf{a}}_{N_a}}(\phi ) = \frac{1}{{\sqrt {N_a} }}{[1,{e^{jkd_a\sin (\phi )}},...,{e^{jkd_a(N - 1)\sin (\phi )}}]^T},
\end{equation}
where $k = 2\pi /\lambda $ and $\lambda$ is the wavelength. $d_a$ is the antenna spacing.
\section{Channel Estimation Procedure}\label{channeles}
In prior works for IRS-assisted systems\cite{huangchi,qinte2,ism,ghy,nby,huangchi2}, perfect CSI of all channels involved is assumed known at every terminal. Unfortunately, owing to the passivity of IRSs, the traditional channel estimation approaches\cite{csi1,csi1a,csi3} are not applicable any more to IRS-assisted systems. Thanks to the natural sparsity of THz channels (see (\ref{sepchan})), techniques of low-complexity beamspace channel estimation, such as beam training\cite{track3,track5}, can be invoked to facilitate the estimation of the multipath components (MPCs) in (\ref{sepchan}). 
\subsection{Quantization Error by Beam Training}\label{co}
Given a propagation path with arbitrary continuous angle ${\varphi ^*}$ in practice, the discrete AoA/AoD of the codewords' beams in channel estimation leads to quantization error. In order to measure the quantization error, the characteristics of the narrow beams are analyzed as follows.

For terminal with $N_a$ antennas, the codeword of normalized narrow beam in direction $\varphi$ is simply given as ${{\bf{a}}_{{N_a}}}(\varphi )$ (see (\ref{ula})). The normalized beam power of ${{\bf{a}}_{{N_a}}}(\varphi )$ in the direction $\psi$ is given as    
\begin{sequation}\label{bpwer}
\begin{split}
A\left( {{{\bf{a}}_{{N_a}}}(\varphi ),\psi } \right) &= \left| {{{\bf{a}}_{{N_a}}}{{(\varphi )}^H}{{\bf{a}}_{{N_a}}}(\psi )} \right|\\
 &= \left| {\frac{1}{{{N_a}}}\sum\limits_{n = 1}^{{N_a}} {{e^{jk{d_a}(n - 1)[\sin (\psi ) - \sin (\varphi )]}}} } \right| \le 1,
\end{split}
\end{sequation}
and we denote the beam coverage of ${{\bf{a}}_{{N_a}}}(\varphi )$ as 
\begin{equation}
{\mathcal{CV}}\left( {{{\bf{a}}_{{N_a}}}(\varphi )} \right) = \left\{ {\psi \left| {\; {A\left( {{{\bf{a}}_{{N_a}}}(\varphi ),\psi } \right)}  \ge \rho } \right.} \right\}
\end{equation}
where $\rho$ is the coverage-edge energy, and the length of coverage is named beam width hereafter. 
\begin{lemma}
Narrow beam in direction $\varphi$ is equivalent to that in direction $\pi-\varphi$ in term of terminal, i.e.,
\begin{equation}
{{\bf{a}}_{{N_a}}}(\varphi)= {{\bf{a}}_{{N_a}}}(\pi-\varphi ).
\end{equation}
Besides, the beam coverage of ${{\bf{a}}_{{N_a}}}(\varphi )$ is front-back mirror  symmetrical about the array plane.
\end{lemma}
\begin{IEEEproof}
It is easy to verify that ${{\bf{a}}_{{N_a}}}(\varphi)= {{\bf{a}}_{{N_a}}}(\pi-\varphi )$. According to (\ref{bpwer}), we get $A\left( {{{\bf{a}}_{{N_a}}}(\varphi ),\psi } \right)=A\left( {{{\bf{a}}_{{N_a}}}(\varphi ),\pi-\psi } \right)$, i.e., ${\mathcal{CV}}\left( {{{\bf{a}}_{{N_a}}}(\varphi )} \right)= \pi-{\mathcal{CV}}\left( {{{\bf{a}}_{{N_a}}}(\varphi )} \right)$. In general, the angle perpendicular to the plane is defined as 0 and the front range is $[-\pi/2,\pi/2]$. Thus, the beam coverage is front-back mirror symmetrical about array plane.
\end{IEEEproof}

According to Lemma 1, beams with front-back mirror symmetry about array plane is isomorphic. For simplicity of exposition hereafterin, we use the same notation $\varphi _i$ for a pair of front-back symmetrical beams. Considering half-wavelength antenna spacing, we have following Propositions and Corollaries for narrow beam design.
\begin{figure}[t]
\center
\includegraphics[width=2.5in]{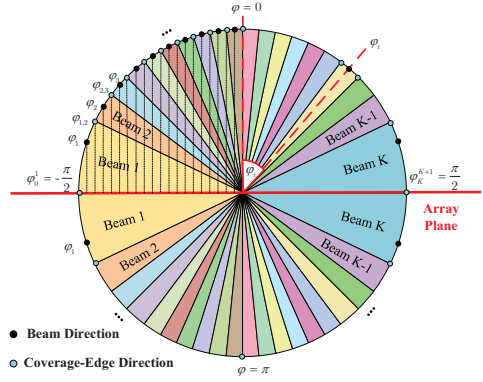}
\caption{Narrow-beam patterns in different directions with common $\rho$.}\label{figpat}
\vspace{-12pt}
\end{figure}
\begin{proposition}
For $K$ ($K\ge N_a$) narrow beams covering all directions with common $\rho$, we use $\{\varphi _i\}_{i=1}^{K}$ to represent the $i$th beam direction and $\{\varphi _i^{i\pm 1}\}_{i=1}^{K}$ to represent the coverage-edge direction of $i$th beam. As shown in Fig. \ref{figpat}, despite the common $\rho$, each beam yields different beam width. For $\forall \;i$th beam, we have 
\begin{equation}
|\sin \varphi _i^{i + 1} - \sin {\varphi _i}| = |\sin {\varphi _i} - \sin \varphi _i^{i - 1}| = \frac{1}{K},
\end{equation}
and 
\begin{equation}
\rho  = \frac{{\sin [({N_a}\pi)/2K]}}{{{N_a}\sin [\pi/2K]}},
\end{equation}
where $\rho$ is monotonically increasing with the increase in $K$ when $K \ge {N_a}$. The beam direction of $\{ {\varphi _i}\} _{i = 1}^{K}$ is given as 
\begin{equation}
{\varphi _i} = \left\{ {\begin{split}
{\arcsin \left[ {\frac{{2i - 1}}{K} - 1} \right],\;\;{\rm{front\;range\;(FR)}}}\\
{\pi  - \arcsin \left[ {\frac{{2i - 1}}{K} - 1} \right],\;\;{\rm{back\;range\;(BR)}}}
\end{split}} \right..
\end{equation}
\end{proposition}
\begin{IEEEproof}
\begin{figure}[t]
\center
\includegraphics[width=2.5in]{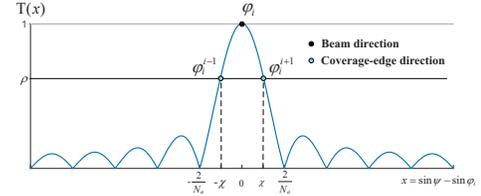}
\caption{Normalized $i$th beam power distribution in different directions.}\label{figsin}
\vspace{-12pt}
\end{figure}
For narrow beam in direction $\varphi$, the normalized beam power in (\ref{bpwer}) can be further expressed as
\begin{align}
A\left( {{{\bf{a}}_{{N_a}}}(\varphi ),\psi } \right) &= \left| {\frac{1}{{{N_a}}}\sum\limits_{n = 1}^{{N_a}} {{e^{jk{d_a}(n - 1)[\sin (\psi ) - \sin (\varphi )]}}} } \right|\notag\\
& =  \left| {\frac{1}{{{N_a}}}\frac{{{e^{j\frac{{{N_a}k{d_a}m}}{2}}}\left( {{e^{j\frac{{{N_a}k{d_a}m}}{2}}} - {e^{ - j\frac{{{N_a}k{d_a}m}}{2}}}} \right)}}{{{e^{j\frac{{k{d_a}m}}{2}}}\left( {{e^{j\frac{{k{d_a}m}}{2}}} - {e^{ - j\frac{{k{d_a}m}}{2}}}} \right)}}} \right|\notag\\
 &= \left| {\frac{1}{{{N_a}}}{e^{j\frac{{({N_a} - 1)k{d_a}m}}{2}}}\frac{{\sin [({N_a}k{d_a}m)/2]}}{{\sin [(k{d_a}m)/2]}}} \right|
\end{align}
where $m = \sin (\psi ) - \sin (\varphi )$. Thus, for half-wavelength antenna spacing, i.e., $d_a=\lambda/2$, the normalized beam power of ${{\bf{a}}_{{N_a}}}(\varphi _ i)$ is given as
\begin{equation}
A\left( {{{\bf{a}}_{{N_a}}}(\varphi _i ),\psi } \right) = \left| {\frac{{\sin [\frac{{{N_a}\pi }}{2}(\sin (\psi ) - \sin (\varphi _i))]}}{{{N_a}\sin [\frac{\pi }{2}(\sin (\psi ) - \sin (\varphi _i))]}}} \right|
\end{equation}
Define a function ${\rm T}(x) = |\sin [\frac{{{N_a}\pi }}{2}x]/\{ {N_a}\sin [\frac{\pi }{2}x]\}|$ as shown in Fig. \ref{figsin}, it can be observed that the beam power monotonically decreases with the increase in $|x|$ when $|x| \leq 2/N_a$. Thus, the common $\rho$ yields common $|x|$ for all beams, and for $\forall \;i$th beam, we have 
\begin{equation}
|\sin \varphi _i^{i + 1} - \sin {\varphi _i}| = |\sin {\varphi _i} - \sin \varphi _i^{i - 1}| = \chi(\rho),
\end{equation}
where $\chi$ is a constant and subjected to the inverse function of $\rho={\rm T}(\chi)$. When all the $K$ beams cover all directions as shown in Fig. \ref{figpat}, we have $2K$ number of $\chi$ uniformly divide the interval [-1,1] with $\chi=1/K$, and the condition $K \ge {N_a} \Rightarrow \chi  <  1/{N_a}$. Thus, $\rho$ monotonically increases with the increase in $K$. In Fig. \ref{figpat}, $\sin {\varphi _1}\;,...,\sin {\varphi _{K}}$ uniformly divide the interval [-1,1] and it is easy to obtain the beam direction by combining the feasible domain of the arcsine function. 
\end{IEEEproof}
\begin{figure}[t]
\center
\includegraphics[width=3.5in]{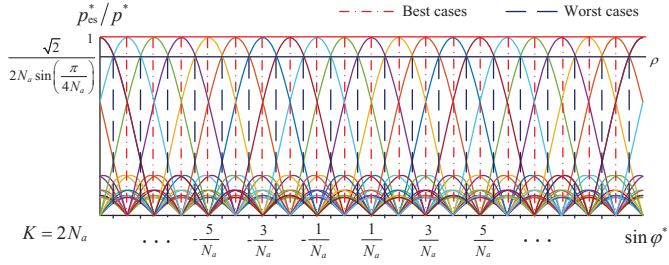}
\caption{Accuracy of measurement versus direction of AoA angle ${\varphi ^*}$.}\label{figscan}
\vspace{-12pt}
\end{figure}
\begin{proposition}
Given a narrow beam with arbitrary AoA angle ${\varphi ^*}$ to terminal with $N_a$ antennas. On condition that $K$ ($K\ge N_a$) narrow beams covering all directions with common $\rho$ are used to estimate the power, the best cases are that ${\varphi ^*}$ lies on the any one of the beam directions and the worst cases are that ${\varphi ^*}$ lies on the any one of the coverage-edge directions. Both cases are presented in Fig. \ref{figscan}, in which $p_{{\rm{es}}}^*/p^*$ is the normalized received energy (see (\ref{ppes}) in appendix).
\end{proposition}
\begin{corollary}
Given a narrow beam with arbitrary AoA angle ${\varphi ^*}$ to terminal with $N_a$ antennas. On condition that $K$ ($K\ge N_a$) narrow beams covering all directions with common $\rho$ are used to estimate the power, the normalized quantization error of worst case is given as
\begin{equation}\label{worst}
{e_{{\rm{worst}}}}=1 - \frac{{\sin [({N_a}\pi )/2K]}}{{{N_a}\sin [\pi /2K]}},
\end{equation}
and the average normalized quantization error is given as 
\begin{align}\label{aver}
{e_{{\rm{aver}}}} =& 1 - \int\limits_{ - 1_+}^{\frac{{2 - K}}{K}} {\frac{{\sin \left[ {\frac{{{N_a}\pi }}{2}\left( {y -\frac{1-K}{K}} \right)} \right]}}{{\sqrt {1 - {y^2}} {N_a}\pi \sin \left[ {\frac{\pi }{2}\left( {y - \frac{1-K}{K}} \right)} \right]}}} {\rm{d}}y \notag\\
-& \sum\limits_{n = 2}^{K-1} {\int\limits_{\frac{{2n - 2 - K}}{K}}^{\frac{{2n - K}}{K}} {\frac{{\sin \left[ {\frac{{{N_a}\pi }}{2}\left( {y - \frac{{2n - 1 - K}}{K}} \right)} \right]}}{{\sqrt {1 - {y^2}} {N_a}\pi \sin \left[ {\frac{\pi }{2}\left( {y - \frac{{2n - 1 - K}}{K}} \right)} \right]}}} {\rm{d}}y} \notag\\
 -& \int\limits_{\frac{{K - 2}}{K}}^{1_-} {\frac{{\sin \left[ {\frac{{{N_a}\pi }}{2}\left( {y - \frac{K-1}{K}} \right)} \right]}}{{\sqrt {1 - {y^2}} {N_a}\pi \sin \left[ {\frac{\pi }{2}\left( {y - \frac{K-1}{K}} \right)} \right]}}} {\rm{d}}y.
\end{align}
\end{corollary}
\begin{IEEEproof}
The proof of Proposition 2 and Corollary 2 are referred to Appendix A. Based on numerical results, we note that a proper $K$ should be $\ge 2N_a$ where ${e_{{\rm{aver}}}}<4\%$.
\end{IEEEproof}

\subsection{Hierarchical Sweeping and Reflecting Modes Realization}
To find the direction with strongest beam power, hierarchical beam sweeping is an efficient way to reduce search complexity without loss optimality. $M$-tree search schemes \cite{track5,track6,csi1,csi1a} realize the beam sweeping stage by stage with decreasing beam range. Above studies assumed the number $K$ of total narrow beams is exactly $K{\rm{ = }}{M^S}$ and the number of stages is given as $S \in \mathbb{N}^+$. Here, we propose a search Criterion for arbitrary $K$. 

\emph{Criterion 1:} For $K$ narrow beams covering all direction, the defined narrow beam is implemented in the bottom stage. Each stage up, the beam implemented covers $M$-fold the range more. As such, the number of total stages is given as $\{{S_M} \in {\mathbb{N}^+ }\;|\log _M^{K} \le {S_M} < \log _M^{K} + 1\} $ and the beam in $s$th stage covers ${M^{S_M - s}}$ narrow beam(s).

\begin{figure}[t]
\center
\includegraphics[width=3.5in]{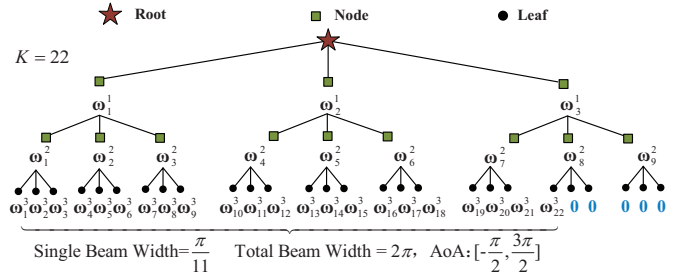}
\caption{The structure of 3-tree hierarchical beam sweeping when $K=22$.}\label{figttr}
\vspace{-6pt}
\end{figure}

Let ${\bm{\omega }}_n^s$ denotes the $n$th beam candidate in the $s$th stage. Considering the arbitrary $K$ which may not exactly be $M^n,\;n \in \mathbb{N}^+$, zero vectors need to be added in the bottom stage for constructing full tree. A example 3-tree structure of such a procedure when $K=22$ is demonstrated in Fig. \ref{figttr}. In each stage, we find and follow the best beam (node) for next-stage search, until the best narrow beam (leaf) is found.

Considering half-wavelength antenna spacing, the narrow beam candidates (leaves) in the bottom stage is exactly given as the receiver's array response vector (\ref{ula}), such as
\begin{equation}\label{phi}
\begin{split}
&{\bm{\omega }}_n^{{S_3}} = {{\bf{a}}_{{N_a}}}({\varphi _n}) = \frac{1}{{\sqrt {{N_a}} }}{[1,{e^{j\pi\sin ({\varphi _n})}},...,{e^{j\pi(N - 1)\sin ({\varphi _n})}}]^T},\\
&{\varphi _n} = \left\{ {\begin{split}
{\arcsin \left( {\frac{{2n - 1}}{K} - 1} \right),\;\;\;{\rm{FR}}}\\
{\pi  - \arcsin \left( {\frac{{2n - 1}}{K} - 1} \right),\;\;{\rm{BR}}}
\end{split}} \right.,\;\;n = 1,...,K.\\
\end{split}
\end{equation}
For wide beam candidates (nodes) design, we first propose the following Criterion for arbitrary $K$.

\emph{Criterion 2:} To design the wide beam candidates, i.e., $\{{\bm{\omega }}_n^s\}_{s=1}^{S_M-1}\in {C^{{N_a} \times 1}}$ in a $M$-tree structure, we expect that 
\begin{equation}\label{cb}
{({\bm{\omega }}_i^{{S_M}})^H}{\bm{\omega }}_n^s = \left\{ {\begin{split}
&{1,\;\;\;{\bm{\omega }}_i^{{S_M}}{\rm{is\;a\;descendant}}\;{\rm{of}}\;{\bm{\omega }}_n^s}\\
&{0,\;\;\;\;\;\;\;\;\;\quad{\rm{otherwise}}}
\end{split}} \right.
\end{equation}
holds true for all $i$ when $n=1,...,S_M-1.$

Based on Criterion 2 for M-tree search, we can write the design of wide beam candidates matrix of $s$th stage, i.e.,
\[{{\bm{\Xi }}_s} = [{\bm{\omega }}_1^s\;{\bm{\omega }}_2^s\;...\;{\bm{\omega }}_{{M^s}}^s]\quad{\rm{for}}\;s = 1,...,{S_M} - 1\]
in a more compact form as 
\begin{equation}\label{com}
{[{\bm{\omega }}_1^{{S_M}}\;{\bm{\omega }}_2^{{S_M}}\;...\;{\bm{\omega }}_{K}^{{S_M}}]^H}{\bm{\Xi }}_s = {{\bf{L}}^H}{\bf{\Xi }} = {{\bf{D}}_s},
\end{equation}
where ${{\bf{D}}_s}$ is an $K \times M^s$ matrix. Define a function as 
\[{\mathcal{U}_{K}}[a] = \left\{ {\begin{split}
{a,\;\;a \le K}\\
{0,\;\;a > K}
\end{split}} \right..\]
The $i$th column of ${{\bf{D}}_s}$ contains element 1 in locations  
\[
\{ {\mathcal{U}_{K}}[(i - 1){M^{{S_M} - s}} + n]\;|\;n = 1,...,{M^{{S_M} - s}}\}, 
\]
and element 0 in other locations. As a result, the wide beam candidate can be projectively given as 
\begin{equation}\label{bems}
{\bm{\omega }}_n^s = \;{({\bf{L}}{{\bf{L}}^H})^{ - 1}}{\bf{L}}{{\bf{D}}_S}(:,n).
\end{equation}
Let $N_s=1$ and $s=1$, the wide beam in hybrid beamforming architecture is realized as \begin{equation}\label{HB}
{\bm{\omega }}_n^s = {\bf{F}}_{RF}{\bf{f}}_B^{s,n}s={\bf{F}}_{RF}{\bf{f}}_B^{s,n}.
\end{equation} 
We state that (\ref{HB}) can be always achieved by using two RF chains, and the solution is given in Proposition 2 of \cite{jingdian}.   
\begin{figure}[t]
\center
\includegraphics[width=3.4in]{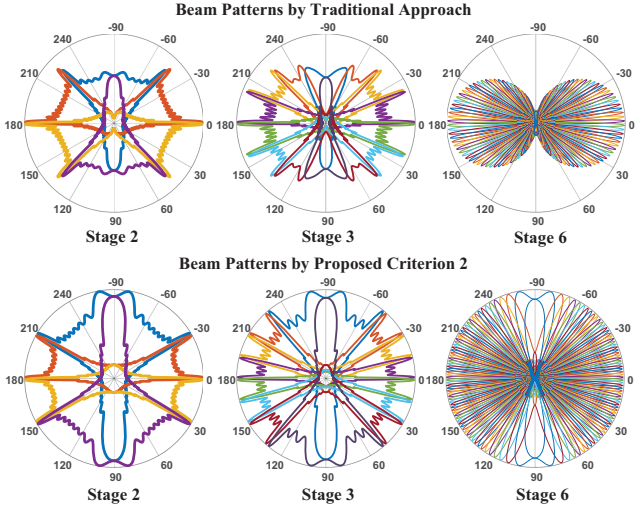}
\caption{Comparison of beam patterns by traditional approach\cite{csi1,csi1a} and proposed Criterion 2 with the setting that $M=2$, $N_a=32$, and $K=64$.}\label{compa}
\vspace{-12pt}
\end{figure}
\begin{remark}
We mention that the proposed Criterion 2 is inspired by prior works \cite{csi1,csi1a} which happen to have the similar concept but contain following shortages.
\begin{itemize}
\item The narrowest beam number must satisfy $K = {M^{n \in {\mathbb{N}^ + }}}$.
\item The derived narrowest beams are with different coverage-edge energy, which may causes direction misalignment.
\end{itemize}

To overcome above shortcomings, we first define the beam with common coverage-edge energy at the bottom stage (see (\ref{phi})). For arbitrary $K$, Criterion 2 is proposed based on leaves in perspective of graph theory (unlike the principle based on ${\bf{a}}_{{N_a}}({\varphi _n})$ in above works). Fig. \ref{compa} illustrates the comparison of beam patterns by different approaches.\end{remark}

For purpose of matching the reflecting state to specific IRS design, we provide the following proposition.
\begin{proposition}
Given an incoming narrow beam ${{\bf{a}}_{{N_r}}}\left( {{\varphi _{{\rm{in}}}}} \right)$ with direction ${\varphi _{{\rm{in}}}}$ to IRS, to achieve a reflecting narrow beam in return path, the phase-shift matrix of \emph{return mode} is given as 
\begin{equation}
\begin{split}
&{\bf{\Theta }}_{\rm{ret}}({\varphi _{{\rm{in}}}}) = {\rm{diag}}(\beta {e^{j\theta _1^{}}},\beta {e^{j\theta _2^{}}}, \cdots ,\beta {e^{j\theta _{{N_r}}^{}}}),\\
&{\theta _n} = -2k{d_a}(n - 1)\sin {\varphi _{{\rm{in}}}},\;\;\;n = 1,...,{N_r}.
\end{split}
\end{equation}
To achieve a reflecting narrow beam in direction ${\varphi _{{\rm{out}}}}$, the phase-shift matrix of \emph{direction mode} is given as 
\begin{align}
&{\bf{\Theta }}_{\rm{dir}}({\varphi _{{\rm{in}}}},{\varphi _{{\rm{out}}}}) = {\rm{diag}}(\beta {e^{j\theta _1^{}}},\beta {e^{j\theta _2^{}}}, \cdots ,\beta {e^{j\theta _{{N_r}}^{}}}),\\
&{\theta _n} = k{d_a}(n - 1)(\sin {\varphi _{{\rm{out}}}}- \sin {\varphi _{{\rm{in}}}}),\;\;\;n = 1,...,{N_r}.\notag
\end{align}
\end{proposition}
\begin{IEEEproof}
The proof are referred to Appendix B, and above defined modes are used in the description below.
\end{IEEEproof}
\subsection{Cooperative Channel Estimation Procedure}
The main parameters of the MPCs in (\ref{sepchan}) include four angles  $\varphi _{A,M}^l,\varphi _{R,M}^l,$ $\varphi _{R,N}^{l},\varphi _{B,N}^{l}$. Owing to the grid effects of channel estimation, we use $\hat{\varphi}$ to represent the estimated direction. To obtain the four angles, two phases are developed to achieve measurements as following. \\
\begin{figure}[t]
\center
\includegraphics[width=3.4in]{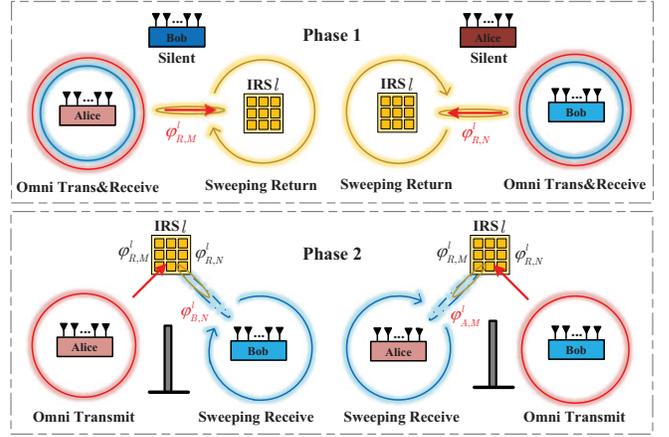}
\caption{Phases of the cooperative channel estimation procedure.}\label{figp2}
\vspace{-12pt}
\end{figure}
\indent \emph{Phase 1}: Keep Bob silent and fix Alice to be concurrently in an omni-beam transmitting and receiving mode. Then, IRS$l$ successively sweeps the narrow beam in return mode, i.e., changing ${\bf{\Theta }}_{\rm{ret}}({\varphi})$, of which the sequence is predefined on time slots and known to all terminals. The best return direction is informed to Alice by determining the time slot with the strongest receiving signal. Repeating operation by switching the manner of Alice and Bob, we obtain $\hat{\varphi} _{R,M}^l$ and $\hat{\varphi} _{R,N}^{l}$.

\indent \emph{Phase 2}: With the obtained $\hat{\varphi} _{R,M}^l$ and $\hat{\varphi} _{R,N}^{l}$, we fix IRS$l$ to optimally bridge the Alice-IRS$l$-Bob link by direction mode, i.e., ${\bf{\Theta }}_{\rm{dir}}(\hat{\varphi} _{R,M}^l,\hat{\varphi} _{R,N}^{l})$. Fix Alice to be in an omni-beam transmitting mode and Bob hierarchically sweeps the beam to find the best direction. Repeating operation by switching the manner of Alice and Bob, we obtain $\hat{\varphi} _{B,N}^l$ and $\hat{\varphi} _{A,M}^{l}$.

Above procedure is illustrated in Fig. \ref{figp2}. It is easy to follow above description for implementation and full algorithms are omitted due to the limited pages. 
\section{Design of IRS and Hybrid Precoder/Combiner}\label{channeles}
With the obtained estimated angles, the optimal design of IRS for communication is given as
\begin{equation}
 {\bf{\Theta }}_l = {\bf{\Theta }}_{\rm{dir}}(\hat{\varphi} _{R,M}^l,\hat{\varphi} _{R,N}^{l}),\;\; l=1,...,N_i.
\end{equation}
By training beams on above angles, i.e., ${\bf{x}} = {\bf{a}}_{{N_r}}^R\left( {\hat\varphi _{B,M}^l} \right)$, ${\bf{w}} = {\bf{a}}_{{N_u}}^U\left( {\hat\varphi _{U,N}^l} \right)$, we can directly obtain the effective composite loss 
\[{a_l} \approx  {\eta _l}G_tG_ra(f,d_l^M)a(f,d_l^N), \;\; l=1,...,N_i,\]
on Alice-IRS$l$-Bob link by calculating the energy of received beam. Next, a key lemma is introduced to provide the basis of the derivations that follows.
\begin{lemma}[{[14]}]
For a ULA system with azimuth angles of arrival or departure drawn independently from a continuous distribution, the transmit and receive array response vectors are orthogonal, i.e., we have ${\bf{a}}({\phi ^k}) \bot {\rm{span}}(\{ {\bf{a}}({\phi ^l})|\forall l \ne k\} )$ as the number of
antenna elements, $N$, tends to infinity and the number of paths in the channel is $L = o(N)$.\end{lemma}

Here, we straightforwardly provide the design of hybrid precoder and combiner as
\begin{align}\label{ds}
{{\bf{F}}_{RF}} &= [{\bf{a}}_{{N_t}}^A{\left( {\hat\varphi _{A,M}^1} \right)},...,{\bf{a}}_{{N_t}}^A{\left( {\varphi _{A,M}^{{N_i}}} \right)},\bf{0},...,\bf{0}],\notag\\
{{\bf{W}}_{RF}} &= [{\bf{a}}_{{N_u}}^B\left( {\varphi _{B,N}^1} \right),...,{\bf{a}}_{{N_u}}^B\left( {\varphi _{B,N}^{{N_i}}} \right),\bf{0},...,\bf{0}],\\
{{\bf{F}}_B} &= {\rm{diag}}(\sqrt{S_1},...,\sqrt{S_{N_i}},0,...,0),\;\;{{\bf{W}}_B} = {{\bf{I}}_{N_{RF}^u}},\notag
\end{align}
where $\{S_i\}_{i=1}^{N_i}$ is the power allocation factors. According to Lemma 2, the transceivers design (\ref{ds}) creates a set of parallel independent reflecting channels and the spectral efficiency is degraded as
\begin{equation}
R = {\log _2}{\rm{det}}\left[ {{{\bf{I}}_{{N_i}}} + \frac{P}{\sigma _n^2}{\bf{F}}_B^H{\rm{diag}}({a_1^2},...,{a_{{N_i}}^2}){{\bf{F}}_B}} \right].
\end{equation}
The remaining objective of Alice is to determine the power factor on these subchannels for maximizing total efficiency, i.e,
\begin{equation}\label{wf}
\begin{split}
&\mathop {\max }\limits_{{S_l}} \log \prod\limits_{l = 1}^{{N_i}} {1 + \frac{P}{\sigma _n^2}{a_l^2}{S_l}} \\
&\;\;{\rm{s}}.{\rm{t}}.\;\;\;\;\sum\nolimits_{l = 1}^{{N_i}} {{S_l}}  = 1,\quad \{ {S_l}\} _{l = 1}^{{N_i}} \ge 0.
\end{split}
\end{equation}
Problem (\ref{wf}) admits a convex form of classic water-filling problem and the optimal solution can be obtained by resorting the Karush-Kuhn-Tucker method, which is given by
\begin{equation}
{S_l} = \left\lceil {\frac{1}{{\ln 2 \cdot \mu }} - \frac{{\sigma _n^2}}{{P{a_l^2}}}} \right\rceil^+ ,\;\;\;l = 1,...,{N_i},
\end{equation}
where ${\left\lceil  \cdot  \right\rceil ^ + } = \max \{\cdot,0\}$. The Lagrange parameter $\mu > 0$ is chosen by 1-D search to satisfy the constraint $\sum\nolimits_{l = 1}^{{N_i}} {{S_l}}=1$.
\section{Simulation and Numerical Results}
\begin{figure}[t]
\center
\includegraphics[width=3.2in]{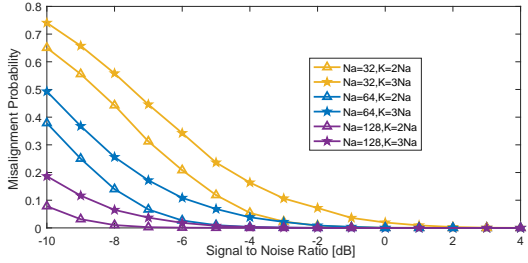}
\caption{Misalignment probability on bottom stage versus signal to noise ratio.}\label{12}
\vspace{-6pt}
\end{figure}
\begin{figure}[t]
\center
\includegraphics[width=2.8in]{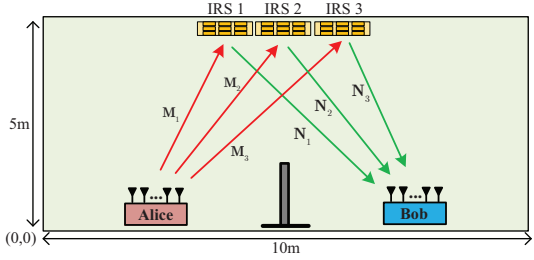}
\caption{The simulated IRS-assisted MIMO communication scenario.}\label{ind}
\vspace{-12pt}
\end{figure}
In this section, we first provide simulation results to illustrate the performance of the proposed training codebook for channel estimation. The results were averaged over 10,000 random channel realizations. Considering a ULA with half-wavelength antenna spacing, Fig. \ref{12} plots the misalignment probability (MP) on bottom stage versus SNR with different implementation parameters. As we can see, the minimum SNR that satisfies zero MP is related to the implementation. It decreases with the increase of antennas number $N_a$ and increases with the increase of narrow beam number $K$.

Then, numerical results are presented to validate the effectiveness of proposed cooperative designs for IRS-assisted massive MIMO system. We consider such a indoor scenario as illustrated in Fig. \ref{ind}, in which Alice and Bob are randomly located at $(0,{d_A} \in [0,5])$ and $(0,{d_B} \in [5,10])$. Three IRS are located at $(5,4)$, $(5,5)$, and $(5,6)$. The results were averaged over 10,000 random location settings of transceivers. In each result that follows, the operating frequency is set as 0.3THz and the background noise power at the receiver is $\sigma _n^2=-80$dBm. The antenna spacing for all terminals is $d_a=\lambda/2$ and the IRS reflection coefficient is $\beta=1$. Absorption coefficient $\tau (0.3T)=0.0033/m$ and antenna gains $G_t=G_r=18$dBi for $N_a=32$ ($G_t=G_r=21$dBi for $N_a=64$). In the following, we respectively plot the spectral efficiency of the receivers in different settings of antenna number and narrow beam number. For comparison, we also plot the upper-bound rate of IRS-assisted scheme and the upper-bound rate of non-IRS-assisted scheme, as two benchmarks. The former is realized by the optimal FDB with the optimal IRSs designs under perfect CSI. The latter one is realized by the optimal FDB with random IRSs setting.
\begin{figure}[t]
\includegraphics[width=3.4in]{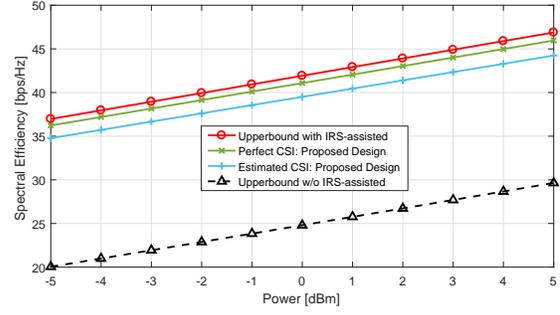}
\caption{Spectral efficiency versus transmit power, where the terminal elements number $N_t=N_i=N_u=32$ and $K=2N_a$.}\label{n32}
\vspace{-10pt}
\end{figure}
\begin{figure}[t]
\includegraphics[width=3.4in]{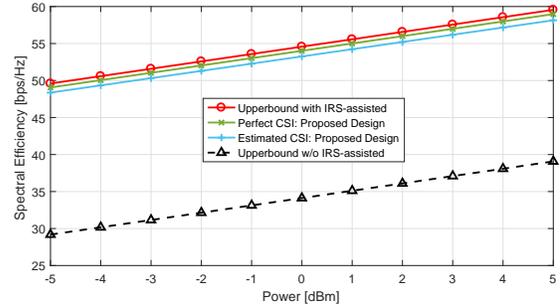}
\caption{Spectral efficiency versus transmit power, where the terminal elements number $N_t=N_i=N_u=64$ and $K=3N_a$.}\label{n64}
\vspace{-12pt}
\end{figure}

Fig. \ref{n32} plots the spectral efficiency versus transmit power, in which $N_t=N_i=N_u=32$ and $K=2N_a$. It is seen that the performance gain of the IRS-assisted schemes is significant compared to the non-IRS-assisted benchmark, since the LoS link is blocked and the reflecting links provide additional strong propagation paths. Moreover, the spectral efficiency achieved by our proposed design with perfect CSI is close to the upperbound, which validates the effectiveness of our proposed solutions. The gap between the performance with perfect CSI and that with estimated CSI is due to the AoAs/AoDs quantization error.

Fig. \ref{n64} plots the spectral efficiency versus transmit power, in which $N_t=N_i=N_u=64$ and $K=3N_a$. Compared to Fig. \ref{n32}, the performance gain of the IRS-assisted schemes in Fig. \ref{n64} increases due to the increase of terminal elements, and the proposed design with perfect CSI is closer to upperbound. In addition, with the increased estimation ratio $K/N_a$ (increased quantization resolution), the performance gap resulted from AoAs/AoDs quantization error is decreasing. 
Thus, with large-scale antenna array implementation and adequate quantization resolution, our proposed channel estimation procedure and cooperative transmission solutions exhibit a comparable performance in the IRS-assisted MIMO system.

\section{Conclusion}
We considered the channel estimation and transmission design in THz massive MIMO IRS-assisted system. Specifically, we invoked the beam training manner to facilitate the channel measurement. Analysis of beam distribution as well as the quantization error by beam training are provided. To reduce the complexity of estimation, a hierarchical codebook design is provided as the basis of beam training. Finally, a cooperative channel estimation procedure is proposed to achieve the measurement of IRS-assisted system. Subsequently, the IRS and hybrid precoder/combiner designs are given in closed form. Numerical results showed significant performance improvement of proposed scheme as compared to non-IRS-assisted one and the achieved spectral efficiency can approach the upperbound benchmark.
    \begin{appendices}
      \section{Proof of Proposition 2}
The quantization error is existing due to the fact that the intended power $p^*$ is not fully estimated by alignment beam, but estimated as 
\begin{equation}\label{ppes}
p_{{\rm{es}}}^* = \mathop {\max }\limits_{n = 1,...,K} {p^*}A\left( {{{\bf{a}}_{{N_a}}}({\varphi ^{\rm{*}}}),{\varphi _n}} \right).
\end{equation}
By Propositon 1, the worst quantization error is given as (\ref{worst}), where we left out the modulus operator because the right-hand side is always positive when $K\ge N_a$.  Considering the arbitrary AoA angle ${\varphi ^*}$ follows a uniform distribution ${\varphi ^*}\sim U[- \pi /2,3\pi /2]$, the probability density function (PDF) of it is given as
\begin{equation}
f_{\Phi}({\varphi ^*}) = \frac{1}{{2\pi }},\;\;\;\;\;{\varphi ^*} \in [ - \pi /2,3\pi ].
\end{equation}
The sine function $y=\sin {\varphi }$ is monotonic at the interval $[ - \pi /2,\pi /2]$ and $[ - \pi /2,3\pi /2]$. The inverse function can be written as ${\varphi _1}(y) = \arcsin (y)$ and ${\varphi _2}(y) = \pi  - \arcsin (y).$
Thus, the PDF of $y=\sin {\varphi }$ is given as 
\begin{equation}\label{pdf}
{f_{\sin \Phi }}(y) = {f_\Phi }[{\varphi _1}(y)]|\varphi _1^{\prime}(y)| + {f_\Phi }[{\varphi _2}(y)]|\varphi _2^{\prime}(y)|.
\end{equation}
The center (resp. coverage edge) angle of the beams 1 to $K$ mapped to $\sin {\varphi}$ is given as $(2n - 1 - K)/K$ (resp. $(2n - 2 - K)/K,\;\;(2n - K)/K$) for $n=1,...,K$. Considering the PDF interval $y \in (-1,1)$ in (\ref{pdf}), the average accuracy is given on three pieces of integral by expectation formula as
\begin{align}\label{acu}
\overline {\frac{{p_{{\rm{es}}}^*}}{{{p^*}}}} &= \int\limits_{ - 1_+}^{\frac{{2 - K}}{K}} {{f_{\sin \Phi }}(y)\frac{{\sin \left[ {\frac{{{N_a}\pi }}{2}\left( {y - \frac{{1 - K}}{K}} \right)} \right]}}{{{N_a}\sin \left[ {\frac{\pi }{2}\left( {y - \frac{{1 - K}}{K}} \right)} \right]}}} {\rm{d}}y\notag\\
 &\quad+ \sum\limits_{n = 2}^{K-1} {\int\limits_{\frac{{2n - 2 - K}}{K}}^{\frac{{2n - K}}{K}} {{f_{\sin \Phi }}(y)\frac{{\sin \left[ {\frac{{{N_a}\pi }}{2}\left( {y - \frac{{2n - 1 - K}}{K}} \right)} \right]}}{{{N_a}\sin \left[ {\frac{\pi }{2}\left( {y - \frac{{2n - 1 - K}}{K}} \right)} \right]}}} {\rm{d}}y} \notag\\
 &\quad+ \int\limits_{\frac{{K - 2}}{K}}^{1_-} {{f_{\sin \Phi }}(y)\frac{{\sin \left[ {\frac{{{N_a}\pi }}{2}\left( {y - \frac{{K-1}}{K}} \right)} \right]}}{{{N_a}\sin \left[ {\frac{\pi }{2}\left( {y - \frac{{K-1}}{K}} \right)} \right]}}} {\rm{d}}y
\end{align}
Substituting (\ref{pdf}) into (\ref{acu}), the average quantization error can be written as (\ref{aver}), which completes the overall proof.

\section{Proof of Proposition 4}
We first derive the IRS solution for direction mode. Given an incoming narrow beam $
{{\bf{a}}_{{N_r}}}\left( {{\varphi _{{\rm{in}}}}} \right)$ with direction ${\varphi _{{\rm{in}}}}$, we expect that the reflecting signal achieves narrow beam in direction ${\varphi _{{\rm{out}}}}$, i.e., 
\begin{align}
&{{\bf{a}}_{{N_r}}}\left( {{\varphi _{\rm{out}}}} \right)\notag \\
&\;= \frac{1}{{\sqrt {{N_r}} }}[1,{e^{jk{d_a}\sin {\varphi _{{\rm{out}}}}}},...,{e^{jk{d_a}({N_r} - 1)\sin {\varphi _{{\rm{out}}}}}}]\notag\\
&\;= {\bf{\Theta }}{{\bf{a}}_{{N_r}}}\left( {{\varphi _{{\rm{in}}}}} \right)\\
&\;= \frac{1}{{\sqrt {{N_r}} }}\left[ {{e^{j{\theta _1}}},{e^{j[k{d_a}\sin {\varphi _{{\rm{in}}}} + {\theta _1}]}},...,{e^{j[k{d_a}({N_r} - 1)\sin {\varphi _{{\rm{in}}}} + {\theta _{{N_r}}}]}}} \right].\notag
\end{align}
Thus, the solution can be derived from the equation as
\begin{equation}
\begin{split}
k{d_a}(n - 1)\sin {\varphi _{{\rm{in}}}} + {\theta _n} = jk{d_a}(n - 1)\sin {\varphi _{{\rm{out}}}}\\
 \Rightarrow \;\;\;{\theta _n} = k{d_a}(n - 1)(\sin {\varphi _{{\rm{out}}}} - \sin {\varphi _{{\rm{in}}}}),\;\forall n.
\end{split}
\end{equation}
Let ${\varphi _{{\rm{out}}}} = \pi  + {\varphi _{{\rm{in}}}}$, we reach the solution of return mode as
\begin{equation}
{\theta _n} = -2k{d_a}(n - 1)\sin {\varphi _{{\rm{in}}}},\;\;\;n = 1,...,{N_r}.
\end{equation}
\end{appendices}


\begin{thebibliography}{1}

\bibitem{IF}
I. F. Akyildiz, J. M. Jornet, and C. Han, ``Terahertz band: Next frontier for wireless communications,'' \emph{Physical Communication (Elsevier),} vol. 12, pp. 16-32, Sep. 2014.

\bibitem{Track}
C. Lin and G. Y. Li, ``Indoor terahertz communications: How many antenna arrays are needed?'' \emph{IEEE Trans. Wireless Commun.,} vol. 14, no. 6, pp. 3097-3107, Jun. 2015.

\bibitem{Jornet}
J. M. Jornet and I. F. Akyildiz, ``Channel modeling and capacity analysis for electromagnetic wireless nanonetworks in the Terahertz band,'' \emph{IEEE Trans. Wireless Commun.}, vol. 10, no. 10, pp. 3211-3221, Oct. 2011.

\bibitem{ZL}
L. Zhang \emph{et al.}, ``Space-time-coding digital metasurfaces,''  \emph{Nat. Commun.,}
vol. 9, pp. 1-11, Oct. 2018.

\bibitem{cometa}
T. J. Cui, M. Q. Qi, X. Wan, J. Zhao, and Q. Cheng, ``Coding metamaterials,
digital metamaterials and programmable metamaterials,'' \emph{Light: Science $\&$ Applications,} vol. 3, no. 10, pp. e218, Oct. 2014.

\bibitem{gain}
W. Tang, M. Z. Chen, X. Chen, J. Y. Dai, Y. Han, M. Di Renzo, Y. Zeng, S. Jin, Q. Cheng, and T. J. Cui, ``Wireless communications with reconfigurable intelligent surface: Path loss modeling and experimental measurement,'' \emph{arXiv:1911.05326,} 2019.
%
%\bibitem{qinte}
%Q. Wu and R. Zhang, ``Intelligent Reflecting Surface Enhanced Wireless Network: Joint Active and Passive Beamforming Design,''  in Proc.\emph{IEEE GLOBECOM}, Abu Dhabi, UAE, pp. 1-6, Dec. 2018,

\bibitem{qinte2} 
Q. Wu and R. Zhang,``Intelligent Reflecting Surface-Enhanced Wireless Network via Joint Active and Passive Beamforming,''  \emph{IEEE Trans. Wireless Commun.}, DOI 10.1109/TWC.2019.2936025, 2019.

\bibitem{huangchi}
C. Huang  \emph{et al.}, ``Achievable rate maximization by passive intelligent mirrors,'' in Proc. \emph{IEEE ICASSP,}  pp. 3714-3718, 2018.

\bibitem{huangchi2}
C. Huang \emph{et al.}, ``Reconfigurable intelligent surfaces for energy efficiency in wireless communication,'' \emph{IEEE Trans. Wireless Commun.,} vol. 18, no. 8, pp. 4157-4170, Aug 2019.

\bibitem{ism}
Q.-U.-A. Nadeem \emph{et al.},``Large intelligent surface assisted MIMO communications,'' \emph{arXiv:1903.08127,} 2019.

\bibitem{ghy}
H. Guo \emph{et al.}, ``Weighted sum-rate optimization for intelligent reflecting surface enhanced wireless networks,'' \emph{preprint arXiv:1905.07920,} 2019.

\bibitem{nby}
B. Ning \emph{et al.}, ``Intelligent Reflecting Surface Design for MIMO System by Maximizing Sum-Path-Gains,'' \emph{preprint arXiv:1909.07282,} 2019.


\bibitem{5-2}
A. Alkhateeb, G. Leus, and R. W. Heath, ``Limited feedback hybrid precoding for multi-user millimeter wave systems,'' \emph{IEEE Trans. Wireless Commun.,} vol. 14, no. 11, pp. 6481-6494, Nov. 2015.

\bibitem{5}
O. El Ayach, R. W. Heath, Jr., S. Abu-Surra, S. Rajagopal, and Z. Pi, ``The capacity optimality of beam steering in large millimeter wave MIMO systems,'' in \emph{Proc. IEEE SPAWC,} pp. 100-104, 2002.

\bibitem{jingdian}
F. Sohrabi and W. Yu, ``Hybrid digital and analog beamforming design for large-scale antenna srrays''  \emph{IEEE J. Sel. Topics Signal Process.}, vol. 10, no. 3, pp. 501-513, Apr. 2016.

\bibitem{csi1}
A. Alkhateeb, O. El Ayach, G. Leus, and R. W. Heath, Jr., ``Channel estimation and hybrid precoding for millimeter wave cellular systems,'' \emph{IEEE J. Sel. Topics Signal Process.,} vol. 8, no. 5, pp. 831-846, Oct. 2014.

\bibitem{csi1a}
M. E. Eltayeb, A. Alkhateeb, R. W. Heath, Jr., and T. Y. Al-Naffouri, ``Opportunistic beam training with hybrid analog/digital codebooks for mmWave systems,'' in \emph{Proc. IEEE Global Conf. Signal Inf. Process. (GlobalSIP)}, Orlando, FL, USA, pp. 315-319, Dec. 2015.

%\bibitem{csi1c}
%C. Liu, M. Li, S. V. Hanly, I. B. Collings and P. Whiting, ``Millimeter wave beam alignment: Large deviations analysis and design insights,'' \emph{IEEE J. Sel. Areas Commun.,}  vol. 35, no. 7, pp. 1619-1631, Jul. 2017


\bibitem{csi3}
Y. Peng, Y. Li, and P.Wang, ``An enhanced channel estimation method for millimeter wave systems with massive antenna arrays,'' \emph{IEEE Commun. Lett.,} vol. 19, no. 9, pp. 1592-1595, Sep. 2015.


\bibitem{track3}
Z. Xiao, L. Bai, and J. Choi, ``Iterative joint beamforming training with constant-amplitude phased arrays in millimeter-wave communications,'' \emph{IEEE Commun. Lett.,} vol. 18, no. 5, pp. 829-832, May 2014.

%\bibitem{track2}
%M. Kokshoorn, P. Wang, Y. Li, and B. Vucetic, ``Fast channel estimation for millimetre wave wireless systems using overlapped beam patterns,'' in \emph{Proc. IEEE Int. Conf. Commun. (ICC),} London, U.K., Jun. 2015, pp. 1304-1309.
%
%\bibitem{track3}
%J. Wang et al., ``Beam codebook based beamforming protocol for multi-Gbps millimeter-wave WPAN systems,'' \emph{IEEE J. Sel. Areas Commun.}, vol. 27, no. 8, pp. 1390-1399, Oct. 2009.
%
%\bibitem{track4}
%S. Hur, T. Kim, D. J. Love, J. V. Krogmeier, T. A. Thomas, and A. Ghosh, ``Millimeter wave beamforming for wireless backhaul and access in small cell networks,'' \emph{IEEE Trans. Commun.}, vol. 61, no. 10, pp. 4391-4403, Oct. 2013.

\bibitem{track5}
Z. Xiao, T. He, P. Xia, and X.-G. Xia, ``Hierarchical codebook design for beamforming training in millimeter-wave communication,'' \emph{IEEE Trans. Wireless Commun.,} vol. 15, no. 5, pp. 3380-3392, May 2016.


\bibitem{track6}
S. Noh, M. D. Zoltowski, and D. J. Love, ``Multi-resolution codebook and adaptive beamforming sequence design for millimeter wave beam alignment,'' \emph{IEEE Trans. Wireless Commun.,} vol. 16, no. 9, pp. 5689-5701, Sep. 2017.




\end{thebibliography}
\end{document}